\documentclass[12pt,titlepage]{article}
\usepackage{bm, amssymb,pifont,cancel, amsmath,comment,color}
\usepackage[dvips]{graphicx}
\makeatletter

\def\slash#1{\not\!\!#1}

\@addtoreset{equation}{section}
\makeatother
 
\begin{document}

\begin{titlepage}
\null
\begin{flushright}
WU-HEP-16-05
\end{flushright}

\vskip 1.5cm
\begin{center}
\baselineskip 0.8cm
{\LARGE \bf
DBI action of real linear superfield in \\4D ${\cal N}=1$ conformal supergravity}

\lineskip .75em
\vskip 1.5cm

\normalsize

{\large Shuntaro Aoki} $\!${\def\thefootnote{\fnsymbol{footnote}}\footnote[1]{E-mail address: shun-soccer@akane.waseda.jp}}, 
{\large and} {\large Yusuke Yamada} $\!${\def\thefootnote{\fnsymbol{footnote}}\footnote[2]{E-mail address: yuusuke-yamada@asagi.waseda.jp}}

\vskip 1.0em

{\small\it Department of Physics, Waseda University, \\ 
Tokyo 169-8555, Japan}

\vspace{12mm}

{\bf Abstract}\\[5mm]
{\parbox{13cm}{\hspace{5mm} \small
%%%%%%%%%%%%%%%%%%%%%%%%%%%%%%%%%%%%%%%%%%%%%%%%%%%%%%%%%%%%%%%%%%%%
We construct the Dirac-Born-Infeld (DBI) action of a real linear multiplet in 4D $\mathcal{N}=1$ supergravity. Based on conformal supergravity, we derive the general condition under which the DBI action can be realized, and show that it can be constructed in the new minimal supergravity. We also generalize it to the matter coupled system.
%%%%%%%%%%%%%%%%%%%%%%%%%%%%%%%%%%%%%%%%%%%%%%%%%%%%%%%%%%%%%%%%%%%%
}}

\end{center}

\end{titlepage}

\tableofcontents
\vspace{35pt}
\hrule
\section{Introduction}
Higher-order derivative terms play important roles in the several contexts, e.g., inflation models, modified gravity, renormalization of gravity, and so on. From a phenomenological and theoretical viewpoint, their embeddings into supersymmetry (SUSY) or supergravity (SUGRA) are also interesting. In particular, there exist many non-renormalizable terms in SUGRA and it is quite natural to consider the extension including higher-order derivative terms and the effects of them on cosmology and particle phenomenology. The higher-order derivative terms of a chiral superfield in 4D SUSY or SUGRA and their cosmological applications have been investigated so far, e.g., in Refs.~\cite{Khoury:2010gb,Khoury:2011da,Baumann:2011nm,Farakos:2012je,Koehn:2012ar,Farakos:2012qu,Koehn:2012te,Farakos:2013zya,Gwyn:2014wna,Aoki:2014pna,Aoki:2015eba,Ciupke:2015msa,Bielleman:2016grv}.

The Dirac-Born-Infeld (DBI) action \cite{Born:1934gh,Dirac:1962iy} includes such higher-order derivative terms. It was first proposed as a nonlinear generalization of Maxwell theory. The DBI action is also motivated by string theory, which is a promising candidate for a unified theory including gravity. In the context of string theory, an effective action of D-brane is described by a DBI-type action, which consists of Maxwell terms $F_{\mu \nu }$ as well as the ones of scalar fields $\partial _{\mu }\phi^i\partial _{\nu }\phi^jg_{ij}$ and a two-form $B_{\mu \nu }$ in general, 
\begin{align}
S_{\rm{DBI}}=\int d^Dx \sqrt{-g}\left( 1-\sqrt{{\rm{det}}(g_{\mu \nu }+ \partial _{\mu }\phi ^i\partial _{\nu }\phi ^jg_{ij}+B_{\mu \nu }+F_{\mu \nu })} \right) . \label{string DBI}
\end{align}

SUSY Dp-brane actions in $D$ dimension are also important for the effective theory of superstring. With a component formalism, such actions have also been discussed in many literature. For example, in Refs.~\cite{Aganagic:1996pe,Aganagic:1996nn}, the authors construct SUSY Dp-brane actions with local kappa symmetry based on a component formalism in 10 dimensional spacetime. In a similar way, the p-brane action in various dimensions has also been discussed in Ref.~\cite{Bergshoeff:2013pia}. In Refs.~\cite{Howe:1996mx,Cederwall:1996pv,Cederwall:1996ri,Bergshoeff:1996tu}, the SUSY Dp-brane in SUGRA background is constructed by considering the background super-vielbein on the brane and couplings between them. 

An approach based on superfields is useful for constructing a manifestly SUSY invariant action and generalizing it. Within the formalism, such 4D $\mathcal{N}=1$ SUSY extensions of the DBI action have been known partially. The DBI action of a vector superfield, which corresponds to the case with $\phi^i=B_{\mu \nu }=0$ in Eq. $\eqref{string DBI}$, is constructed in Refs.~\cite{Cecotti:1986gb,Bagger:1996wp,Rocek:1997hi,Kuzenko:2002vk,Kuzenko:2005wh}. In particular, in Refs.~\cite{Bagger:1996wp,Rocek:1997hi}, it is shown that such an action appears from the partial breaking of 4D $\mathcal{N}=2$ SUSY. Its SUGRA embedding has also been discussed in Refs.~\cite{Cecotti:1986gb,Kuzenko:2002vk,Kuzenko:2005wh,Abe:2015nxa}. Its application to inflation models has been investigated in Ref.~\cite{Abe:2015fha}. Furthermore, in global SUSY,  multiple $U(1)$~\cite{Ferrara:2014oka,Ferrara:2014nwa} and massive~\cite{Ferrara:2015ixa}  extensions of the DBI action have been discussed. In particular, for the case with multiple U(1) vector multiplets, linear actions ~\cite{Andrianopoli:2014mia}, general conditions for partial SUSY breaking ~\cite{Andrianopoli:2015wqa,Andrianopoli:2015rpa}, and c-maps ~\cite{Andrianopoli:2016eub} have also been discussed.
 
For the DBI action of scalar fields, which corresponds to the case with $F_{\mu \nu }=B_{\mu \nu }=0$ in Eq. $\eqref{string DBI}$, its SUSY extension has been done via partially broken $\mathcal{N}=2$ SUSY theory, where the Goldstino multiplet is an $\mathcal{N}=1$ real linear superfield \cite{Rocek:1997hi,Bagger:1997pi,GonzalezRey:1998kh}. However, there has never been the SUGRA extension of the DBI action of a real linear superfield. In this paper, we discuss the embedding of the DBI action of a real linear superfield into SUGRA. The action of a chiral superfield can be found in Ref.~\cite{Koehn:2012ar}. In general, it is known that the action with a chiral superfield can be rewritten in terms of the one with a real linear superfield, and vice versa (via linear-chiral duality \cite{Siegel:1979ai}). Therefore, our action, which will be discussed in this paper, would be equivalent to that derived in Ref.~\cite{Koehn:2012ar} through the duality transformation. We will discuss this point and the differences between their result and ours.

In Refs. \cite{Rocek:1997hi,Bagger:1997pi,GonzalezRey:1998kh}, the DBI action of a real linear multiplet is realized with a chiral multiplet, which is constrained by a specific $\mathcal{N}=1$ SUSY constraint. We will investigate the corresponding constraint which is a key for the construction of DBI action, in SUGRA. To achieve this, we use a formulation based on conformal SUGRA \cite{Kaku:1978nz,Kugo:1982cu,Kugo:1983mv}\footnote{We will use the superconformal tensor calculus~\cite{Kaku:1978nz,Kugo:1982cu,Kugo:1983mv}. See also another formulation, conformal superspace~\cite{Butter:2009cp,Kugo:2016zzf}.}, where  one can treat off-shell SUGRA with different sets of auxiliary fields in a unified manner. Because of the restrictions on the SUGRA embedding of the ${\cal N}=1$ constraint, we will find that the DBI action of a real linear superfield can be realized only in the so-called new minimal formulation of SUGRA. Furthermore, we will extend the DBI action to the matter coupled version of it.

The remaining parts of this paper are organized as follows. First, we will briefly review the SUSY DBI action of a real linear superfield in Sec.~\ref{review}. There, we will find that the constraint imposed between a chiral and real linear superfield is important for the construction. Then, we will extend the constraint to that in conformal SUGRA in Sec.~\ref{extension}. After a short review of conformal SUGRA, we will also review the concept of the {\it{u-associated}} derivative which is crucial for the superconformal extension. Using this {\it{u-associated}} derivative, we will complete the embedding and find that the constraint can be consistently realized in the new minimal SUGRA. With the constraint, we will construct the corresponding action in the new minimal SUGRA, and write down the bosonic component action in Sec.~\ref{Component action}. The linear -chiral duality and the matter coupled extension will be also discussed there. Finally, we will discuss the correspondence and differences between results in related works and ours in Sec.~\ref{discussion}, and summarize this paper in Sec.~\ref{summary}. In Appendix.~\ref{explicit}, the explicit components of the multiplet including the {\it{u-associated}} derivative are shown.

In this paper, we use the unit $M_P=1$ where $M_P=2.4\times 10^{18}$ GeV is the reduced Planck mass, and follow the conventions of \cite{Wess:1992cp} in Sec.~\ref{review} and of \cite{Freedman:2012zz} in other parts. $a,b\cdots$ denote Minkowski indices and $\mu,\nu\cdots$ denote curved indices.

%%%%%%%%%%%%%%%%%%%%%%%%%%%%%%%%%%%%%%%%%%%%%%%%%%%%%%%%%%%%%%%%%%%%%%%%%%%%%%%%%%%%%%%%%%%%%%%%%%%%%%%%%%%%%%%%%%%%%%%%%%%%%%%%%%%
\section{Review of DBI action in global SUSY}\label{review}
In this section, we briefly review the DBI action of a real linear superfield in global SUSY \cite{Bagger:1997pi}.  We use a chiral superfield $X$ and a real linear superfield $L$ which satisfy the conditions,
\begin{align}
\bar{D}_{\dot{\alpha }} X=0, \ \ \ D^2L=\bar{D}^2L=0,
\end{align}
where $D_{\alpha }$ and $\bar{D}_{\dot{\alpha }}$ are a SUSY spinor derivative and its complex conjugate.
To construct the DBI action for $L$, we consider the following constraint between $X$ and $L$,
\begin{align}
X-\frac{1}{4}X\bar{D}^2\bar{X}-\bar{D}_{\dot{\alpha }}L\bar{D}^{\dot{\alpha }}L=0 , \label{global constraint}
\end{align}
where $\bar{X}$ is a complex conjugate of $X$ \footnote{In Ref.~\cite{Bagger:1997pi}, the constraint $\eqref{global constraint}$ has been obtained from the tensor multiplet in $\mathcal{N}=2$ SUSY through partial breaking of it. Here, we do not discuss its origin and we just use the constraint as a guideline to obtain the DBI action. In Sec.~\ref{discussion}, we will briefly comment on the relation between the partial breaking of ${\cal N}=2$ SUSY and our construction.}. The equation. $\eqref{global constraint}$ can be solved with respect to $X$ and we obtain
\begin{align}
X=\bar{D}_{\dot{\alpha }}L\bar{D}^{\dot{\alpha }}L +\frac{1}{2}\bar{D}^2\Biggl[ \frac{D^{\alpha }LD_{\alpha }L\bar{D}_{\dot{\alpha }}L\bar{D}^{\dot{\alpha }}L }{1-\frac{1}{2}A+\sqrt{1-A+\frac{1}{4}B^2}}\Biggr],\label{solution for global constraint}
\end{align}
where
\begin{align}
A\equiv \frac{1}{2}\{ D^2( \bar{D}_{\dot{\alpha }}L\bar{D}^{\dot{\alpha }}L)+{\rm{h.c.}}  \}, \ \ \ B\equiv \frac{1}{2}\{ D^2( \bar{D}_{\dot{\alpha }}L\bar{D}^{\dot{\alpha }}L)-{\rm{h.c.}}  \}.
\end{align}
Using this solution $\eqref{solution for global constraint}$, we can construct the SUSY DBI action as 
\begin{align}
\mathcal{L}=\int d^2 \theta X(L) +{\rm{h.c.}}. \label{global DBI}
\end{align}
One can check that the bosonic part of the Lagrangian $\eqref{global DBI}$ produces,
\begin{align}
\mathcal{L}_B=1-\sqrt{1-B\cdot B+\partial C\cdot \partial C-(B\cdot \partial C)^2} ,\label{component form of global DBI}
\end{align}
where $C$ and $B_a$ are a real scalar and a constrained vector satisfying $\partial ^a B_a=0$, in the real linear superfield, and we use the notation $B\cdot \partial C \equiv B^a \partial _aC $. It is known that, through the linear-chiral duality, Eq. $\eqref{component form of global DBI}$ produces the DBI action of a complex scalar, which can be interpreted as the 4D effective D3-brane action. 
We call Eq. $\eqref{component form of global DBI}$ the DBI action of a real linear superfield in this paper.

It is worth noting that Eq. $\eqref{solution for global constraint}$ satisfies the nilpotency condition, i.e., $X^2=0$, due to the Grassmann property of the SUSY spinor derivative, $\bar{D}_{\dot{\alpha }}$. This reflects the underlying Volkov-Akulov SUSY \cite{Volkov:1972jx,Rocek:1978nb}. Instead of writing the action like Eq. $\eqref{global DBI}$, we can also rewrite the same system imposing the constraint $\eqref{global constraint}$ by a chiral superfield Lagrange multiplier $\Lambda $,
\begin{align}
\mathcal{L}=\int d^2 \theta \biggl[ X +\Lambda \left( X-\frac{1}{4}X\bar{D}^2\bar{X}-\bar{D}_{\dot{\alpha }}L\bar{D}^{\dot{\alpha }}L\right) +\tilde{\Lambda }X^2\biggr]  +{\rm{h.c.}}. \label{global DBI with constraint}
\end{align}
Here we have introduced another Lagrange multiplier $\tilde{\Lambda }$, which ensures the nilpotency of $X$. Indeed, we need not require this condition in the Lagrangian since $X$ satisfies $X^2=0$ after integrating out $\Lambda$ first and solving $X$ with respect to $L$, but the condition is still consistent and makes the calculation simple as far as we focus on the bosonic part of the action, as we will see in the following section.
%%%%%%%%%%%%%%%%%%%%%%%%%%%%%%%%%%%%%%%%%%%%%%%%%%%%%%%%%%%%%%%%%%%%%%%%%%%%%%%%%%%%%%%%%%%%%%%%%%%%%%%%%%%%%%%%%%%%%%%%%%%%%%%%%%%
\section{Extension to 4D $\mathcal{N}=1$ conformal SUGRA}\label{extension}
In this section, we generalize the SUSY DBI action $\eqref{global DBI with constraint}$ discussed in Sec.~\ref{review} to that in SUGRA.  

\subsection{Review of conformal SUGRA} \label{conformal SUGRA}
To construct the action in SUGRA, we use conformal SUGRA formulation. Then, let us briefly review the basics of the conformal SUGRA before proceeding to the specific construction of the DBI action. 

In this formulation, there are extra gauge symmetries such as dilatation, $U(1)_A$ symmetry, S-SUSY and conformal boost in addition to translation, Lorentz transformation and SUSY. The commutation and anti-commutation relations are governed by the superconformal algebra and its representation $\Phi$ called a superconformal multiplet has the following components, 
\begin{align}
\Phi =\{ \mathcal{C}, \mathcal{Z},\mathcal{H},\mathcal{K},\mathcal{B}_a,\Lambda ,\mathcal{D}\} ,\label{general multiplet}
\end{align}
where $\mathcal{Z}$ and $\Lambda $ are spinors; $\mathcal{B}_a$ is a vector; the others are complex scalars.  We also denote the superconformal multiplet $\Phi$ by its first component $\mathcal{C}$,
\begin{align}
\Phi =\langle \mathcal{C} \rangle ,
\end{align}
where $\langle  ...\rangle $ represents the superconformal multiplet which has $\mathcal{C}$ as the first component. $\mathcal{C}$ must be invariant under the transformations of S-SUSY and conformal boost in order for $\Phi=\langle \mathcal{C}\rangle$ to be a superconformal multiplet \cite{Kugo:1983mv}.
 
A superconformal multiplet is characterized by the charge $(w,n)$ under dilatation and $U(1)_A$ symmetry called the Weyl weight and the chiral weight, respectively. For example, a chiral multiplet $X$ has $(w,w)$, in order to satisfy
\begin{align}
\bar{\mathcal{D}}_{\dot{\alpha}}X=0, \label{dbar}
\end{align}
where $\bar{\mathcal{D}}_{\dot{\alpha}}$ is a spinor derivative \cite{Kugo:1983mv}. For a real linear multiplet $L$ defined by, 
\begin{align}
\Sigma L =\bar{\Sigma } L=0,
\end{align}
where $\Sigma $ ($\bar{\Sigma }$) is a (anti-) chiral projection operator, the values of each weight are determined as $(w,n)=(2,0)$. We will discuss these operators, $\mathcal{D}_\alpha$ and $\Sigma $, more precisely in the following subsections.

The chiral multiplet consists of the following components, $\{z,P_L\chi,F\}$, where $z$ and $F$ are complex scalars and $P_L\chi$ is a chiral spinor; $P_L=(1+\gamma_5)/2$ is a left-handed projection operator. It is embedded into a general superconformal multiplet $\eqref{general multiplet}$ as
\begin{align}
\{ z,-\sqrt{2}iP_L\chi , -F,iF,iD_az,0,0  \} , \label{embedding chiral}
\end{align}
where $D_a$ is a superconformal covariant derivative. On the other hand, a real linear multiplet has components, $\{C,Z,B_a\}$, where $C$ is a real scalar, $Z$ is a Majorana spinor and $B_a$ is a constrained vector which satisfies $D^aB_a=0$. A real linear multiplet is embedded into a general superconformal multiplet $\eqref{general multiplet}$ as
\begin{align}
\{ C,Z,0,0,B_a,-\slash{D}Z,-\Box C\}, \label{embedding linear}
\end{align}
where $\slash{D} \equiv \gamma ^a D_a$.

For later convenience, we also introduce a multiplication rule for superconformal multiplets. For a function of multiplets $f(\mathcal{C}^I)$, where $I$ classifies different multiplets, we have
\begin{align}
\nonumber \langle f({\cal C}^I)\rangle=\biggl[ &f,f_I\mathcal{Z}^I, f_I\mathcal{H}^I-\frac{1}{4}f_{IJ}\bar{\mathcal{Z}}^J\mathcal{Z}^I, f_I\mathcal{K}^I+\frac{i}{4}f_{IJ}\bar{\mathcal{Z}}^J\gamma _5\mathcal{Z}^I, f_I\mathcal{B}_a^I-\frac{i}{4}f_{IJ}\bar{\mathcal{Z}}^J\gamma_a \gamma _5\mathcal{Z}^I,\\
\nonumber &f_I\Lambda^I-\frac{i}{2}\gamma _5\left( \mathcal{K}^I-\slash{\mathcal{B}}^I-i\gamma _5\slash{D}\mathcal{C}^I+i\gamma _5\mathcal{H}^I\right) f_{IJ}\mathcal{Z}^J-\frac{1}{4}\left( \bar{\mathcal{Z}}^J\mathcal{Z}^I\right) \mathcal{Z}^Kf_{IJK},\\
\nonumber &f_I\mathcal{D}^ I+\frac{1}{2}f_{IJ}\left( \mathcal{K}^I\mathcal{K}^J+\mathcal{H}^I\mathcal{H}^J-\mathcal{B}^{aI}\mathcal{B}_a^J-D_a\mathcal{C}^ID^a\mathcal{C}^J-2\bar{\mathcal{Z}}^J\Lambda^I-\bar{\mathcal{Z}}^J\slash{D}\mathcal{Z}^I\right)\\
&-\frac{1}{4}f_{IJK}\bar{\mathcal{Z}}^J(\mathcal{H}^K-i\gamma _5\mathcal{K}^K-i\slash{\mathcal{B}}^K\gamma _5)\mathcal{Z}^I+\frac{1}{16}f_{IJKL}(\bar{\mathcal{Z}}^J\mathcal{Z}^I)(\bar{\mathcal{Z}}^K\mathcal{Z}^L) \biggr] , \label{formulaF}
\end{align}
where $f_{IJ\cdots }$ is $\partial f/\partial \mathcal{C}^I\partial\mathcal{C}^J\cdots$ and $\bar{\mathcal{Z}}\equiv \mathcal{Z}^T\hat{C}$ ($\hat{C}$ is a charge conjugation matrix).

We also need action formulas to construct a superconformal action. For a chiral multiplet $X=\{z,P_L\chi,F\}$ with its weight $(3,3)$, there exists the so-called F-term formula~\cite{Kugo:1982cu}, 
\begin{align}
[X]_F=\int d ^{4}x\sqrt{-g}{\rm{Re}} \biggl[ F+\frac{1}{\sqrt{2}}  \bar {\psi }_{\mu}\gamma ^{\mu}P _{L}\chi +\frac{1}{2}z\bar {\psi }_{\mu}\gamma ^{\mu \nu}P _{R}\psi _{\nu} \biggr] , \label{Fformula}
\end{align}
where $\psi _{\mu}$ is a gravitino.
For a real multiplet $\phi =\{C,Z,H,K,B_a,\Lambda,D\}$ with its weight $(2,0)$, we can apply the following D-term formula~\cite{Kugo:1982cu},
\begin{align}
\nonumber  [\phi ]_D= \int d^{4}x\sqrt{-g}\biggl[ &D-\frac{1}{2}i\bar{\psi}\cdot \gamma \gamma _{5}\lambda  -\frac{1}{3}CR+\frac{1}{3}(C\bar{\psi }_{\mu}\gamma ^{\mu \rho \sigma }-i\bar{Z }\gamma ^{\rho \sigma }\gamma _{5})D_{\rho }\psi _{\sigma }\\
    &+\frac{1}{4} \varepsilon ^{abcd}\bar {\psi }_{a}\gamma _{b}\psi _{c}\left(B _{d}-\frac{1}{2}\bar{\psi }_{d}Z \right)\biggr] . \label{Dformula}
\end{align}
Here, all the components of $\phi$ are real (Majorana).

Using these superconformal multiplets, the multiplication rule $\eqref{formulaF}$, and the action formulas $\eqref{Fformula}$ and $\eqref{Dformula}$, we can construct superconformal invariant actions. Finally, we fix some parts of the extra gauge symmetries by imposing the condition to one of the superconformal multiplets $\Phi_0$ called a compensator multiplet, and obtain the $\rm{Poincar\acute{e}}$ SUGRA action.

\subsection{{\it{u-associated}} derivative} \label{u-associated derivative}
Now, we have prepared the tool for constructing the DBI action in SUGRA. Within the conformal SUGRA formulation, we will discuss a constraint corresponding to that in global SUSY, 
\begin{align}
X-\frac{1}{4}X\bar{D}^2\bar{X}-\bar{D}_{\dot{\alpha }}L\bar{D}^{\dot{\alpha }}L=0 , \label{global constraint 2}
\end{align}
in the following. However, it seems to be a nontrivial task to extend the term including SUSY spinor derivatives,
\begin{align}
\bar{D}_{\dot{\alpha }}L\bar{D}^{\dot{\alpha }}L \label{global derivative}
\end{align}
 to that in conformal SUGRA. 
 
To treat the term $\eqref{global derivative}$ in conformal SUGRA, we need the spinor derivative defined as a superconformal operation. In Ref.~\cite{Kugo:1983mv}, it is pointed out that the spinor derivative in conformal SUGRA, $\mathcal{D}_{\alpha }$ ($\bar{\mathcal{D}}_{\dot{\alpha }}$), cannot be defined on a superconformal multiplet $\Phi$ unless $\Phi$ satisfies a specific weight condition, $w=-n$ ($w =n$). This is because $\mathcal{D}_{\alpha }\Phi $ ($\bar{\mathcal{D}}_{\dot{\alpha }}\Phi$) is not generically a superconformal multiplet, i.e., the first component of it is S-SUSY and conformal boost inert only when $w=-n$ ($w=n$) is satisfied. 
Then, it is obvious that we cannot define $\bar{\cal D}_{\dot{\alpha}}L$ as a superconformal multiplet since $L$ has the weight with $(2,0)$.

However, the authors in Ref.~\cite{Kugo:1983mv} also proposed an improved spinor derivative operation, which can be defined on any supermultiplet. They introduced another multiplet, ${\bf{u}}$, called a {\it{u-associated}} multiplet, 
\begin{align}
{\bf{u}} =\{ \mathcal{C}_u, \mathcal{Z}_u,\mathcal{H}_u,\mathcal{K}_u,\mathcal{B}_{au},\Lambda _u,\mathcal{D}_u \} ,
\end{align}
in order to force the first component of ${\cal D}_\alpha \Phi$ to be invariant under S-SUSY and conformal boost. To be specific, they defined the {\it{u-associated}} spinor derivative as 
\begin{align}
\mathcal{D}^{({\bf{u}})}_{\alpha }\Phi =\langle (P_L\mathcal{Z})_{\alpha }+i(n+w)\lambda _{\alpha }\mathcal{C}\rangle ,\ \ \ \lambda _{\alpha } \equiv \frac{i(P_L\mathcal{Z}_u)_{\alpha }}{(w_u+n_u)\mathcal{C}_u}, \label{def u}
\end{align}
where $w_u$ and $n_u$ are the Weyl and chiral weight of a {\it{u-associated}} multiplets, respectively. Unless $w_u+n_u=0$, we can choose any multiplet as the {\it{u-associated}} multiplet. 
%Indeed, ...  
Then, we can define the spinor derivative for an arbitrary superconformal multiplet by this {\it{u-associated}} spinor derivative.

For our purpose, we need the {\it{u-associated}} spinor derivative acting on a real linear multiplet, $\mathcal{D}^{({\bf{u}})}_{\alpha }L $. More generally, we can consider 
\begin{align}
\mathcal{D}^{({\bf{u}}_1)}_{\alpha }({\bf{u}}_2L), \label{u-derivative}
\end{align}
where ${\bf{u}}_1$ is a {\it{u-associated}} multiplet and ${\bf{u}}_2$ is an additional multiplet. These multiplets must satisfy ${\bf{u}}_1\neq {\bf{u}}_2$, since $\mathcal{D}^{({\bf{u}})}_{\alpha }{\bf{u}}$ is identically zero obviously from the definition $\eqref{def u}$.\footnote{As we will discuss, we choose ${\bf{u}}_1$ and ${\bf{u}}_2$ as compensators, which become some parts of the gravity multiplet after superconformal gauge fixings. In the global SUSY expression $\eqref{global derivative}$, all the fields in the gravitational multiplet decouple from it. Therefore, it is natural to consider a possibility that a compensator appears as in Eq. $\eqref{u-derivative}$.} Using this {\it{u-associated}} spinor derivative, Eq. $\eqref{global derivative}$ can be generalized to the one in conformal SUGRA as
\begin{align}
\frac{1}{{\bf{u}}_3}\bar{\mathcal{D}}^{({\bf{u}}_1)}(\bar{\bf{u}}_2L)\bar{\mathcal{D}}^{({\bf{u}}_1)}(\bar{\bf{u}}_2L) , \label{u-derivative part}
\end{align}
where we have introduced a new multiplet ${\bf{u}}_3$\footnote{We will refer all of ${\bf {u}}_i$ as {\it u-associated} multiplets.} for generality and omitted the spinor index, $\dot{\alpha }$, and we have also defined the conjugate of a {\it{u-associated}} derivative as $\bar{\mathcal{D}}_{\dot{\alpha }}^{\bf{u}} \Phi = (\mathcal{D}_{\alpha }^{\bf{u}}(\Phi )^*)^*$ following Ref. \cite{Kugo:1983mv}.

Let us comment on the weight of the multiplet $\eqref{u-derivative part}$. The operator $\bar{\mathcal{D}}^{({\bf{u}})}_{\dot{\alpha} }$ has the weight $(1/2,3/2)$, then the total weight of Eq. $\eqref{u-derivative part}$ is $(2w_2-w_3+5,2n_2-n_3+3)$, where $w_i$ and $n_i$ with $i=1,2,3$ are the Weyl and chiral weights of ${\bf{u}}_i$, respectively. 

Furthermore, Eq. $\eqref{global constraint 2}$ is a ``chiral" constraint since the first and second term in Eq. $\eqref{global constraint 2}$ are chiral multiplets. Then, we require a condition that the multiplet $\eqref{u-derivative part}$ is a chiral multiplet, that is, 
\begin{align}
\bar{\mathcal{D}}\biggl[ \frac{1}{{\bf{u_3}}}\bar{\mathcal{D}}^{({\bf{u_1}})}(\bar{\bf{u_2}}L)\bar{\mathcal{D}}^{({\bf{u_1}})}(\bar{\bf{u_2}}L)  \biggr]  =0 . \label{condition for chiral}
\end{align}
To apply $\bar{\mathcal{D}}$ for Eq. $\eqref{u-derivative part}$, the Weyl and chiral weight of Eq. $\eqref{u-derivative part}$ must satisfy $w=n$ as mentioned before,
\begin{align}
2w_2-w_3+5=2n_2-n_3+3. \label{weight condition between w3n3 and w2n2}
\end{align}
The condition $\eqref{condition for chiral}$ implies that 
\begin{align}
P_R \mathcal{Z}'=0, \label{condition for chiral2}
\end{align}
where $P_R= (1-\gamma _5)/2$ is a right-handed projection operator and $\mathcal{Z}'$ is the second component of the multiplet $\eqref{u-derivative part}$. The equation $\eqref{condition for chiral2}$ can be written explicitly as
\begin{align}
\nonumber &\bar{\tilde{\mathcal{Z}}}_2^cP_R\tilde{\mathcal{Z}}_2^c\biggl[P_R\tilde{Z}+k   P_R\tilde{\mathcal{Z}}_1^c- P_R\tilde{\mathcal{Z}}_3\biggr] 
+\bar{\tilde{Z}}P_R\tilde{Z}\biggl[P_R\tilde{\mathcal{Z}}_2^c+k P_R\tilde{\mathcal{Z}}_1^c- P_R\tilde{\mathcal{Z}}_3\biggr] \\
\nonumber &-k  \bar{\tilde{\mathcal{Z}}}_1^cP_R\tilde{\mathcal{Z}}_1^c\biggl[\left( 1-2k \right) \left( P_R\tilde{Z}+P_R\tilde{\mathcal{Z}}_2^c\right) +P_R\tilde{\mathcal{Z}}_3\biggr] \\
\nonumber &-2k \biggl[\bar{\tilde{\mathcal{Z}}}_2^cP_R\tilde{\mathcal{Z}}_1^c \left( 2P_R\tilde{Z}-P_R\tilde{\mathcal{Z}}_3\right)+\bar{\tilde{Z}}P_R\tilde{\mathcal{Z}}_1^c\left( 2P_R\tilde{\mathcal{Z}}_2^c-P_R\tilde{\mathcal{Z}}_3\right) \biggr] \\
&-2i\biggl[i\tilde{\mathcal{H}}_2^*+\tilde{\mathcal{K}}_2^*-k \left( i\tilde{\mathcal{H}}_1^*+\tilde{\mathcal{K}}_1^*\right) \biggr] \biggl[P_R\tilde{\mathcal{Z}}_2^c+P_R\tilde{Z}-k P_R\tilde{\mathcal{Z}}_1^c\biggr] 
-2\bar{\tilde{\mathcal{Z}}}_2^cP_R\tilde{Z}P_R\tilde{\mathcal{Z}}_3=0,\label{explicit condotion for chiral} 
\end{align}
where 
\begin{align}
&{\bf{u}}_i =\{ \mathcal{C}_i, \mathcal{Z}_i,\mathcal{H}_i,\mathcal{K}_i,\mathcal{B}_{ai},\Lambda _i,\mathcal{D}_i \} ,\ \ \ (i=1,2,3),\\
&\tilde{Z}\equiv \frac{1}{C}Z, \ \ \ \tilde{\mathcal{Z}}_i\equiv \frac{1}{C_i}\mathcal{Z}_i, \ \ \ \tilde{\mathcal{H}}_i(\tilde{\mathcal{K}}_i)\equiv \frac{1}{C_i}\mathcal{H}_i(\mathcal{K}_i), \label{tilde}\\
&k \equiv \frac{w_2+n_2+2}{w_1+n_1},
\end{align}
and $``c"$ denotes the charge conjugation for spinors. 

As a summary, we find that the superconformal realization of Eq. $\eqref{global derivative}$ is the multiplet $\eqref{u-derivative part}$ satisfying the conditions $\eqref{weight condition between w3n3 and w2n2}$ and $\eqref{explicit condotion for chiral}$. 

\subsection{Old minimal versus New minimal}
We have found, in the previous subsection ~\ref{u-associated derivative}, the conditions for extending Eq. $\eqref{global derivative}$ to that in conformal SUGRA. Here, we will choose a conformal compensator $\Phi_0$ as {\it{u-associated}} multiplets, ${\bf{u}}_i$. Then, we have two choices of compensators; one of them is a chiral compensator $S_0$ realizing the old minimal SUGRA and the other is a real linear compensator $L_0$ realizing the new minimal SUGRA.\footnote{We do not discuss the case of the   non-minimal formulation which is realized with a complex linear compensator.}  

Now, we will examine what forms of ${{\bf{u}}_i}$ with both compensators are allowed. Let us start from the old minimal SUGRA realized with a chiral compensator,
\begin{align}
S_0=\{ z_0,-\sqrt{2}iP_L\chi_0 , -F_0,iF_0,iD_az_0,0,0  \},
\end{align}
with its weight $(1,1)$. Here we assume that the multiplets ${\bf{u}}_i$ take the following form 
\begin{align}
{\bf{u}}_i=S_0^{p_i}\bar{S}_0^{q_i},\ \ \ (i=1,2,3), \label{s0s0}
\end{align}
where $p_i$ and $q_i$ are the power of $S_0$ and $\bar{S}_0$, and satisfy $p_1\neq 0$ since $w_1+n_1=(p_1+q_1)+(p_1-q_1)=2p_1$ must be nonzero by a definition of the {\it{u-associated}} multiplet. Here we have to stress that Eq. $\eqref{s0s0}$ is the most general form except for the case including derivative operators on a compensator,\footnote{For example, $S_0\Sigma \bar{S}_0$ could be considered.} which might produce higher-derivative terms of gravity. Using Eq. $\eqref{embedding chiral}$ and the multiplication rule $\eqref{formulaF}$, the components of the multiplet in Eq. $\eqref{s0s0}$ are written as 
\begin{align}
\nonumber &\{ \mathcal{C}_i, \mathcal{Z}_i,\mathcal{H}_i,\mathcal{K}_i,\mathcal{B}_{ai},\Lambda _i,\mathcal{D}_i \} \\
\nonumber  &= \{ z_0^{p_i}z_0^{*q_i},\sqrt{2}iz_0^{p_i-1}z_0^{*q_i-1}(q_iz_0P_R\chi_0 -p_iz_0^*P_L\chi_0 ),\\
\nonumber &z_0^{p_i-2}z_0^{*q_i-2}\left( -q_iz_0^2z_0^*F_0^*-p_iz_0z_0^{*2}F_0+\frac{1}{2}q_i(q_i-1)z_0^2\bar{\chi}_0P_R\chi _0+\frac{1}{2}p_i(p_i-1)z_0^{*2}\bar{\chi}_0P_L\chi _0\right) ,\\
\nonumber &z_0^{p_i-2}z_0^{*q_i-2}\left( -iq_iz_0^2z_0^*F_0^*+ip_iz_0z_0^{*2}F_0+\frac{i}{2}q_i(q_i-1)z_0^2\bar{\chi}_0P_R\chi _0-\frac{i}{2}p_i(p_i-1)z_0^{*2}\bar{\chi}_0P_L\chi _0\right) , \\
&...,...,...\}, \label{explicit s0s0}
\end{align}
where we have omitted the components, $\mathcal{B}_{ai},\Lambda _i$ and $\mathcal{D}_i $, which are not necessary to evaluate  Eq. $\eqref{explicit condotion for chiral}$.  
One finds that Eq. $\eqref{explicit condotion for chiral}$ cannot be satisfied by Eq. $\eqref{s0s0}$ by the following reason: Terms including ${\cal H}_i$ and ${\cal K}_i$ must vanish by themselves since any other terms cannot cancel them. After substituting Eq. $\eqref{explicit s0s0}$ into such a part, we obtain
\begin{align}
\nonumber &i\mathcal{\tilde{H}}_2^*+\tilde{\mathcal{K}}_2^*-k \left( i\tilde{\mathcal{H}}_1^*+\tilde{\mathcal{K}}_1^*\right) =2iF_0^*z_0^{*-1}+i\bar{\chi}_0P_R\chi _0z_0^{*-2}(p_2^2-p_2p_1-p_1+1).
\end{align}
Apparently, the first term cannot be eliminated no matter how we choose the parameters $p_i$ and $q_i$, and the other terms in Eq. $\eqref{explicit condotion for chiral}$ cannot eliminate it because they do not contain $F_0^*$. 
Therefore, we find that Eq. $\eqref{s0s0}$ cannot be a solution of Eq. $\eqref{explicit condotion for chiral}$. This means that Eq. $\eqref{u-derivative part}$ cannot be realized as a chiral constraint in the old minimal SUGRA.

Next, we examine the case in the new minimal SUGRA with a real linear compensator
\begin{align}
L_0=\{ C_0,Z_0,0,0,B_{0a},-\slash{D}Z_0,-\Box C_0\}
\end{align}
with its weight $(2,0)$. In the same way as the old minimal case, we assume the general form of ${\bf{u}}_i$ as
\begin{align}
{\bf{u}}_i=L_0^{r_i},\ \ \ (i=1,2,3), \label{l0}
\end{align}
whose components are 
\begin{align}
\nonumber &\{ \mathcal{C}_i, \mathcal{Z}_i,\mathcal{H}_i,\mathcal{K}_i,\mathcal{B}_{ai},\Lambda _i,\mathcal{D}_i \} \\
 &= \{ C_0^{r_i}, r_iC_0^{r_i-1}Z_0,-\frac{1}{4}r_i(r_i-1)C_0^{r_i-2}\bar{Z}_0Z_0, \frac{i}{4}r_i(r_i-1)C_0^{r_i-2}\bar{Z}_0\gamma _5Z_0,...,...,...\} . \label{explicit l0}
\end{align}
Here we have used Eq. $\eqref{embedding linear}$ and Eq. $\eqref{formulaF}$. Then, after substituting Eq. $\eqref{explicit l0}$ into Eq. $\eqref{explicit condotion for chiral}$ with the Fierz rearrangement, Eq. $\eqref{explicit condotion for chiral}$ is summarized as
\begin{align}
(2r_2-r_3+1)\left\{ CP_RZ\bar{Z}_0P_RZ_0+C_0P_RZ_0\bar{Z}P_RZ\right\}=0. \label{L0 condition}
\end{align}
To satisfy Eq. $\eqref{L0 condition}$, the coefficient must be zero,
\begin{align}
2r_2-r_3+1=0.\label{L0 condition2}
\end{align}
Then, we find that the chiral condition $\eqref{explicit condotion for chiral}$ is satisfied as long as the {\it{u-associated}} multiplets follow the condition $\eqref{L0 condition2}$.

Noting that $w_i=2r_i$ and $n_i=0$ in the ansatz $\eqref{l0}$, the weight condition $\eqref{weight condition between w3n3 and w2n2}$ which the chiral multiplet should obey is now reduced to
\begin{align}
2r_2-r_3+1=0. \label{L0 condition22}
\end{align}
This is nothing but Eq. $\eqref{L0 condition2}$ which is satisfied automatically.

Therefore, we conclude that one can make a multiplet in Eq. $\eqref{u-derivative part}$ a chiral one with the real linear compensator if Eq. $\eqref{L0 condition2}$ is satisfied. Here and hereafter, we focus on the case of the new minimal SUGRA with $r_1=r_3=1$ and $r_2=0$ for simplicity. In this case, the multiplet in Eq. $\eqref{u-derivative part}$ becomes
\begin{align}
\frac{1}{L_0}\bar{\mathcal{D}}^{(L_0)}L\bar{\mathcal{D}}^{(L_0)}L . \label{Lo derivative part}
\end{align}
We present the components of this chiral multiplet $\eqref{Lo derivative part}$ explicitly in Appendix A.

\subsection{Embedding the constraint into conformal SUGRA} \label{embedding}
Let us consider the remaining terms, $X$ and $X\bar{D}^2\bar{X}$ in Eq. $\eqref{global constraint 2}$. For $X$, we just regard it as a superconformal chiral multiplet with the weight $(w,w)$. In order to extend the second one, $X\bar{D}^2\bar{X}$, to a superconformal multiplet, we replace it with $X\Sigma \bar{X}$, where $\Sigma $ is a chiral projection operator in conformal SUGRA. However, $\Sigma $ cannot always be applied for any multiplet $\Phi$ in the same way as the spinor derivative $\mathcal{D}$. It can be applied only when $\Phi$ satisfies the following weight condition, 
\begin{align}
w_{\Phi }=n_{\Phi }+2. \label{condition for projection}
\end{align}
Therefore, we compensate the weight of $\bar{X}$, which has the weight $(w,-w)$, by the real linear compensator multiplet $L_0^s$, where $s$ is the power of $L_0$, 
\begin{align}
X\Sigma \left( \frac{1}{L_0^s}\bar{X} \right) . \label{compensated second term}
\end{align}
Here, the term, $\frac{1}{L_0^s}\bar{X}$, has the weight $(-2s+w,-w)$. According to Eq. $\eqref{condition for projection}$, $s$ must satisfy the condition,
\begin{align}
s=w-1. \label{weight condition of v1}
\end{align}
Taking into account this condition and the fact that $\Sigma $ raises the weight by (1,3), Eq. $\eqref{compensated second term}$ has the weight $(3,3)$, which is correct  for a chiral multiplet. Since the total weight of Eq. $\eqref{compensated second term}$ must be the same as the first term $X$, the value of $w$ is determined as 
\begin{align}
w=3. \label{weight condition of v2}
\end{align}
Then, we find $s=2$ from Eq.  $\eqref{weight condition of v1}$, and Eq. $\eqref{compensated second term}$ becomes  
\begin{align}
X\Sigma \left( \frac{1}{L_0^2}\bar{X} \right) . \label{compensated second term2}
\end{align}
Finally, the weight of the multiplet in Eq. $\eqref{u-derivative part}$ with that in Eq. $\eqref{l0}$ is $(3,3)$ as long as Eq. $\eqref{L0 condition22}$ is satisfied, then Eq. $\eqref{Lo derivative part}$ is automatically satisfied. 

Therefore, we find the complete embedding of a global SUSY expression $\eqref{global constraint 2}$,
\begin{align}
X+\frac{1}{2}X\Sigma \left( \frac{1}{L_0^2}\bar{X} \right) +\frac{1}{4L_0}\bar{\mathcal{D}}^{(L_0)}L\bar{\mathcal{D}}^{(L_0)}L =0,\label{local constraint}
\end{align}
where $X$ is a chiral multiplet with $(3,3)$, $L$ is a real linear multiplet with $(2,0)$, and $L_0$ is a real linear compensator with $(2,0)$.

%%%%%%%%%%%%%%%%%%%%%%%%%%%%%%%%%%%%%%%%%%%%%%%%%%%%%%%%%%%%%%%%%%%%%%%%%%%%%%%%%%%%%%%%%%%%%%%%%%%%%%%%%%%%%%%%%%%%%%%%%%%%%%%%%%%
\section{Component action}\label{Component action}
In this section, we derive the DBI action based on the constraint $\eqref{local constraint}$ in the new minimal SUGRA.

\subsection{Minimal action} \label{Minimal action}
We first consider the minimal extension of the action $\eqref{component form of global DBI}$. The action corresponding to Eq. $\eqref{global DBI with constraint}$ is expected to be
\begin{align}
S=&[2X]_F+\Biggl[2\Lambda \left\{ X+\frac{1}{2}X\Sigma \left(  \frac{\bar{X}}{L_0^2}\right) +\frac{1}{4L_0}\bar{\mathcal{D}}^{(L_0)}L\bar{\mathcal{D}}^{(L_0)}L \right\} \Biggr] _F+[\tilde{\Lambda }X^2]_F +\Biggl[\frac{3}{2}L_0V_R\Biggr] _D, \label{SL0}
\end{align}
where $V_R\equiv \log \frac{L_0}{S\bar{S}}$, $S$ is a chiral multiplet with $(1,1)$, and we have assigned the weights of the Lagrange multiplier chiral multiplet $\Lambda$ to $(0,0)$ and also $\tilde{\Lambda }$ to $(-3,-3)$ in such a way that the total weight is equal to $(3,3)$.
The last term in Eq. $\eqref{SL0}$ is responsible for the kinetic term of the gravitational multiplet. Note that this term is invariant under the transformation $S\to Se^{i\Theta}$ where $\Theta$ is a chiral multiplet with the weight $(0,0)$ since $[L_0(\Theta+\bar{\Theta})]_D\equiv0$ by the nature of a real linear multiplet. Due to this additional gauge invariance, we have gauge degrees of freedom other than superconformal ones. After imposing the gauge fixing condition for this additional gauge symmetry as $S=\{1,0,0\}$, the bosonic part of $\eqref{SL0}$ is given by
\begin{align}
\nonumber S_B=& \int d^4 x \sqrt{-g}\Biggl[ \Biggl( F_X (1+\Lambda )-\frac{|F_X|^2\Lambda }{C_0^2}-\frac{\Lambda}{4C_0}(B_a-i\hat{D}_aC)^2\\
\nonumber &+\frac{C\Lambda}{2C_0^2}(B_a-i\hat{D}_aC)(B_0^a-i\hat{D}^aC_0)
-\frac{C^2\Lambda}{4C_0^3}(B_{0a}-i\hat{D}_aC_0)^2 +{\rm{h.c.}}\Biggr) \\
&-\frac{3}{2}\hat{\Box} C_0\log C_0-\frac{3}{2}\hat{\Box} C_0-\frac{3}{4C_0}(B_0\cdot B_0+\hat{D}C_0\cdot \hat{D}C_0) 
+3A\cdot B_0 \Biggr] , \label{SBfull}
\end{align}
where $\Lambda$ and $F_X$ are a scalar component of the chiral multiplet $\Lambda$ and an auxiliary field of $X$, and $\hat{D}_{\mu}$ is a superconformal covariant derivative only including bosonic fields, for example,
\begin{align}
\hat{D}_{\mu}C=\partial_{\mu}C-2b_{\mu}C,
\end{align}
where $b_{\mu}$ is the gauge field of dilatation. The third term in Eq. $\eqref{SL0}$, $\tilde{\Lambda }X^2$, imposes the nilpotency condition for $X$. Thanks to this, we can drop the scalar component of the chiral multiplet $X$ since the first scalar component can be represented as a fermion bilinear after solving $X^2=0$. That is why, we have inserted this term into the action from the beginning. 
Integrating out the gauge field of $U(1)_A$ symmetry $A_\mu$, we obtain    
\begin{align}
B_{0a}=0.
\end{align}
To eliminate the dilatation symmetry and conformal boost symmetry, we impose the following $D$-gauge and $K$-gauge conditions, 
\begin{align}
C_0=1,\ \ \ \ b_{\mu}=0.
\end{align}
These conditions simplify the action $\eqref{SBfull}$, which becomes
\begin{align}
\nonumber S_B=\int d^4 x \sqrt{-g}\Biggl[ &\frac{1}{2}R+\Bigl( F_X (1+\Lambda ) -|F_X|^2\Lambda \\
&-\frac{\Lambda }{4}(B\cdot B-2iB\cdot \partial C-\partial C\cdot \partial C)+{\rm{h.c.}}\Bigr) \Biggr] . \label{FX}
\end{align}
Then, eliminating the auxiliary field $F_X$ leads to
\begin{align}
S_B=\int d^4 x \sqrt{-g}\Biggl[ &\frac{1}{2}R+\frac{1}{2\lambda }\Bigl(  (\lambda +1)^2+\chi ^2\Bigr) -\frac{1}{2}(B\cdot B-\partial C\cdot \partial C)\lambda -B\cdot \partial C \chi \Biggr] , \label{FXX}
\end{align}
where $\lambda={\rm Re}\Lambda$ and $\chi={\rm Im}\Lambda$.
Finally, we obtain the following conditions from the E.O.Ms for $\lambda $ and $\chi $,
\begin{align}
&\frac{\chi }{\lambda }=B\cdot \partial C ,\\
&\frac{1}{\lambda ^2}=1-(B\cdot \partial C)^2-B\cdot B+\partial C\cdot \partial C .
\end{align}
Substituting them into the action $\eqref{FXX}$, we obtain the on-shell DBI action of a real linear multiplet,
\begin{align}
 S_B=\int d^4 x \sqrt{-g}\Biggl[ \frac{1}{2}R+1-\sqrt{1-B\cdot B+\partial C\cdot \partial C-(B\cdot \partial C)^2}\Biggr] .  \label{DBI}
\end{align}
This is almost the same form as Eq. $\eqref{component form of global DBI}$ except for that our action $\eqref{DBI}$ is formulated in curved background.

Before closing this subsection, let us discuss the linear-chiral duality. It is known that the action of a real linear multiplet can be rewritten in terms of that of a chiral multiplet. However, in the case with the action including derivative terms such as Eq. $\eqref{SL0}$, it is nontrivial to take this duality transformation in a manifestly SUSY way.\footnote{In global SUSY, the dual action has been obtained at the level of superfield in Ref.~\cite{GonzalezRey:1998kh}.} Then, we focus only on the bosonic part $\eqref{DBI}$ and discuss this duality at the component level of bosonic part.

We start from the following Lagrangian which is the relevant part in the action $\eqref{DBI}$,
\begin{align}
\mathcal{L}=1-\sqrt{1-B\cdot B+\partial C\cdot \partial C-(B\cdot \partial C)^2}. \label{Ori}
\end{align}
To rewrite this Lagrangian $\eqref{Ori}$ in terms of the complex scalar of a chiral multiplet, we first relax the constraint on the vector field $B_a$. We impose it by the E.O.M for a scalar field $\ell$, that is, we use 
\begin{align}
\mathcal{L}=1-\sqrt{1-B\cdot B+\partial C\cdot \partial C-(B\cdot \partial C)^2}+B\cdot \partial \ell ,\label{LL}
\end{align}
where $B_a$ is now an unconstrained vector. The Lagrangian $\eqref{LL}$ is equivalent to the original one $\eqref{Ori}$ since the variation with respect to $\ell$ leads to the constraint, $\partial _a B^a=0$.
Instead of $\ell$, varying with respect to $B_a$ gives
\begin{align}
\partial ^a\ell + (\partial ^a C B\cdot \partial C+B^a) \{1-B\cdot B+\partial C\cdot \partial C-(B\cdot \partial C)^2\} ^{-1/2} =0. \label{du}
\end{align}
Our task is now to solve this equation $\eqref{du}$ with respect to $B_a$.
By taking scalar products of Eq. $\eqref{du}$ with $B_a, \partial _a C$ and $\partial _a \ell $, we obtain three independent equations and can solve them with respect to $B^2$, $B\cdot\partial C$, and $B\cdot \partial \ell$. The solutions are
\begin{align}
&B^2=\frac{(\partial \ell) ^2(1+(\partial C)^2)^2-(\partial C\cdot \partial \ell )^2(2+(\partial C)^2)}{Y^2},\\
&B\cdot \partial C=-\frac{\partial C\cdot \partial \ell }{Y},\\
&B\cdot \partial \ell =\frac{-(\partial \ell) ^2(1+(\partial C)^2)+(\partial C\cdot \partial \ell )^2}{Y},
\end{align}
where
\begin{align}
Y \equiv \{ (1+(\partial C)^2)(1+(\partial \ell) ^2)-(\partial C\cdot \partial \ell )^2\}^{1/2}.
\end{align}
Substituting these solutions into the Lagrangian $\eqref{LL}$, we obtain the dual action,
\begin{align}
\nonumber \mathcal{L}&=1-\sqrt{1+(\partial C)^2+(\partial \ell) ^2+(\partial C)^2(\partial \ell) ^2-(\partial C\cdot \partial \ell )^2}\\
&=1-\sqrt{1+\partial \phi \cdot \partial \bar{ \phi }-\frac{1}{4}(\partial \phi)^2 (\partial \bar{ \phi })^2+\frac{1}{4}(\partial \phi \cdot \partial \bar{ \phi })^2} ,\label{LC}
\end{align}
where we have defined a complex scalar $\phi =\ell + iC$. The Lagrangian $\eqref{LC}$ can be written as the DBI form
\begin{align}
\mathcal{L}=1-\sqrt{{\rm{det}} \left( g_{ab}+\frac{1}{2}\partial_a \phi \partial_b \bar{\phi} \right)}.\label{LC2}
\end{align}
This Lagrangian $\eqref{LC2}$ agrees with the one constructed in Ref.~\cite{Koehn:2012ar} using a chiral multiplet directly.

\subsection{Matter coupled extension} \label{matter}
Finally, we discuss the matter coupled DBI action given by
\begin{align}
S=&[2f(\Phi^{I})X]_F+\left[2\Lambda \left\{ X+\frac{1}{2}X\Sigma\left(\frac{\bar{X}}{M(L_0,\Phi^I,\bar{\Phi}^{\bar{J}})}\right)+\frac{1}{4L_0}\bar{\cal D}^{(L_0)}L\bar{\cal D}^{(L_0)}L\right\}\right]_F\nonumber\\
&+[{\cal F}(L_0,\Phi^I,\bar{\Phi}^{\bar{J}})]_D+[\tilde{\Lambda}X^2]_F\label{mDBI},
\end{align}
where $\Phi^I$ ($\bar{\Phi}^{\bar{J}}$) is a (anti-) chiral matter multiplet; $f(\Phi)$ is a holomorphic function of $\Phi^I$ with $(0,0)$; $M(L_0,\Phi^I,\bar{\Phi}^{\bar{J}})$ and ${\cal F}(L_0,\Phi^I,\bar{\Phi}^{\bar{J}})$ are real functions of $\Phi^I,\bar{\Phi}^{\bar{J}}$ and  $L_0$ with $(4,0)$ and $(2,0)$, respectively. Note that we have omitted superpotential term $[W(\Phi^I)]_F$, where $W(\Phi^I)$ is a holomorphic function of $\Phi^I$ with the weight $(w,n)=(3,3)$, since the term is irrelevant to the following discussion. Taking into account the nilpotency condition on $X$, the bosonic component of the action~(\ref{mDBI}) is given by
\begin{align}
S_B=&\int d^4x \sqrt{-g}\Biggl[ \Biggl( F_X(f+\Lambda) -\frac{\Lambda |F_X|^2}{M}-\frac{\Lambda}{4C_0}(B_a-i\hat{D}_aC)^2 \nonumber\\
&+\frac{C\Lambda}{2C_0^2}(B_a-i\hat{D}_aC)(B_0^a-i\hat{D}^aC_0)-\frac{C^2\Lambda}{4C_0^3}(B_0^a-i\hat{D}^aC_0)^2+{\rm h.c.}\Biggr)+{\cal L}_m\Biggr],
\end{align}
where
\begin{align}
{\cal L}_m=&-\frac{1}{3}({\cal F}-{\cal F}_{C_0}C_0)R(b)+\frac{1}{2}{\cal F}_{C_0C_0}(\hat{D}C_0\cdot \hat{D}C_0-B_0\cdot B_0)\nonumber\\
&+2{\cal F}_{I\bar{J}}(F^I\bar{F}^{\bar{J}}-\hat{D}\Phi^I \cdot \hat{D}\bar{\Phi}^{\bar{J}})+\left(-i{\cal F}_{C_0I}B_0\cdot \hat{D}\Phi^I+{\rm h.c.}\right)\label{defLm}.
\end{align}
In the above expression, $\Phi^I$ ($\bar{\Phi}^{\bar{J}}$) and $F^I$ ($\bar{F}^{\bar{J}}$) represent the scalar and auxiliary components of the (anti-) chiral matter multiplet, and subscripts denote the derivative with respect to the corresponding scalar. $R(b)$ becomes a Ricci scalar when $b_{\mu}=0$ is imposed as the $K$-gauge condition.

Before setting superconformal gauge conditions, we integrate out the auxiliary field $F_X$ and the Lagrange multiplier $\Lambda$. We can easily solve the E.O.M for $F_X$ and obtain
\begin{align}
S_B=&\int d^4x \sqrt{-g}\Biggl[\frac{M}{2\lambda}\left\{(\lambda+p)^2+(\chi+q)^2\right\}-\frac{\lambda}{2C_0}(B\cdot B-\hat{D}C\cdot \hat{D}C)\nonumber\\
&-\frac{\chi}{C_0}B \cdot \hat{D}C+\frac{C\lambda}{C_0^2}(B_0\cdot B-\hat{D}C_0\cdot \hat{D}C)+\frac{C\chi}{C_0^2}(B_0\cdot \hat{D}C+B\cdot \hat{D}C_0)\nonumber\\
&-\frac{C^2\lambda}{2C_0^3}(B_0 \cdot B_0-\hat{D}C_0\cdot \hat{D}C_0)-\frac{C^2\chi}{C_0^3}B_0\cdot \hat{D}C_0+{\cal L}_m\Biggr],\label{mDBI2}
\end{align}
where $\lambda={\rm Re}\Lambda$, $\chi={\rm Im}\Lambda$, $p={\rm Re}f$, and $q={\rm Im}f$. Note that, at this stage, the matter Lagrangian ${\cal L}_m$ is not affected by the DBI sector. Next, we eliminate $\lambda$ and $\chi$ by using their E.O.Ms, which are given by
\begin{align}
&-\frac{M}{2\lambda^2}\left\{(\lambda+p)^2+(\chi+q)^2\right\}+\frac{M}{\lambda}(\lambda+p)+\mathcal{A}=0,\\
&\frac{M}{\lambda}(\chi+q)+\mathcal{B}=0,
\end{align}
where
\begin{align}
&\mathcal{A}\equiv -\frac{1}{2C_0}(B\cdot B-\hat{D}C\cdot \hat{D}C)+\frac{C}{C_0^2}(B_0\cdot B-\hat{D}C_0\cdot \hat{D}C)-\frac{C^2}{2C_0^3}(B_0\cdot B_0-\hat{D}C_0\cdot \hat{D}C_0),\\
&\mathcal{B}\equiv -\frac{1}{C_0}B\cdot \hat{D}C+\frac{C}{C_0^2}(B_0\cdot \hat{D}C+B\cdot \hat{D}C_0)-\frac{C^2}{C_0^3}B_0\cdot \hat{D}C_0.
\end{align}
Solutions for them are
\begin{align}
&\lambda|_{\rm sol}^{-1}=\frac{1}{p}\sqrt{1+\frac{2\mathcal{A}}{M}-\frac{\mathcal{B}^2}{M^2}},\\
&\chi|_{\rm sol}=-q-\frac{\lambda|_{\rm sol}}{M}\mathcal{B}.
\end{align} 
Substituting the above solutions into the action~(\ref{mDBI2}), we obtain a relatively simple form
\begin{align}
S_B=\int d^4x\sqrt{-g}\left[Mp\left(1-\sqrt{1+\frac{2\mathcal{A}}{M}-\frac{\mathcal{B}^2}{M^2}}\right)-q\mathcal{B}+{\cal L}_m\right].\label{mDBI3}
\end{align}

The remaining issue is the elimination of auxiliary fields $B_0^a$ and $A_a$. However, it is difficult to do it because of the presence of nonlinear terms of $B_0^a$ contained in the first term in Eq.~(\ref{mDBI3}). In addition, ${\cal L}_m$ has $A_aA^a$ as well as mixing terms between $B_0^a$ and $A_a$ in general cases. Therefore, integration of those auxiliary fields is technically difficult and we cannot obtain the complete on-shell action.\footnote{The general matter coupled system in the new minimal SUGRA not including higher-order derivative terms can be found in Ref.~\cite{Ferrara:1983dh}.}

Although a general case is difficult to complete the remaining task, we can continue our discussion for the following special case. Let us consider the following choice of ${\cal F}(L_0,\Phi^I,\bar{\Phi}^{\bar{J}})$,
\begin{align}
{\cal F}=L_0\log \left(\frac{L_0G(\Phi^i,\bar{\Phi}^{\bar{j}})}{S\bar{S}}\right), \label{special form}
\end{align}
where $\Phi^i$ is a matter chiral multiplet with its weight $(0,0)$, $G(\Phi^i,\bar{\Phi}^{\bar{j}})$ is a real function of $\Phi^i$ and $\bar{\Phi}^{\bar{j}}$, and $S$ is a chiral multiplet with $(1,1)$. This action is also invariant under the transformation $S\to Se^{i\Theta}$ in the same way as the last term in Eq. $\eqref{SL0}$, which characterizes the new minimal SUGRA.

We use the $D$-gauge condition to make the Ricci scalar term canonical. From Eq.~(\ref{defLm}), we can find an appropriate $D$-gauge choice~\cite{Ferrara:1983dh}
\begin{align}
{\cal F}-{\cal F}_{C_0}C_0=-\frac{3}{2}.
\end{align}
As the choice of the additional gauge, we set ${\cal F}_{C_0}=0$~\cite{Ferrara:1983dh}. Then, we can solve these gauge conditions with respect to $C_0$ and $S$ and obtain
\begin{align}
S\bar{S}=&\frac{3}{2}eG,\\
C_0=&\frac{3}{2}.
\end{align} 
Using the $K$-gauge, we also set a condition $b_\mu=0$.

Under these conditions, ${\cal L}_m$ becomes
\begin{align}
{\cal L}_m=&\frac{1}{2}R+2{\cal F}_{i\bar{j}}(F^i\bar{F}^{\bar{j}}-\partial_a\Phi^i\partial^a\bar{\Phi}^{\bar{j}})-\frac{1}{2}B_0^aB_{0a}\nonumber\\
&+(-i{\cal F}_{C_0i}B_0^a\partial_a \Phi^i+{\rm h.c.})+(iB_0^a\partial_a\log S+{\rm h.c.})+2B_0^aA_a,
\end{align}
where $A_a$ is the $U(1)_A$ gauge field mentioned above. We find that the E.O.M for $A_a$ gives a constraint $B_0^a=0$ and the difficulty due to the nonlinear term of $B_0^a$ is circumvented in this case. This result is irrelevant to other parts of the action (\ref{mDBI3}) since they do not contain terms of $A_a$. $F^i$ can be eliminated by their E.O.Ms, and we finally obtain the following on-shell action,
\begin{align}
S_B=\int d^4x\sqrt{-g}\left[Mp\left(1-\sqrt{1+\frac{2\mathcal{A}}{M}-\frac{\mathcal{B}^2}{M^2}}\right)-q\mathcal{B}+\frac{1}{2}R-2{\cal F}_{i\bar{j}}\partial_a\Phi^i\partial^a\bar{\Phi}^{\bar{j}}\right], \label{special matter}
\end{align}
with
\begin{align}
\mathcal{A}=\frac{1}{3}(\partial C\cdot \partial C-B\cdot B),\ \ \ \mathcal{B}=-\frac{2}{3}B\cdot \partial C.
\end{align}
Here, the real function $M$ should be understood as $M|_{C_0=3/2}$.
Note that, in this case, we cannot add superpotential terms of $\Phi^i$ by the following reason: To obtain the constraint $B_0^a=0$, we assumed that only $S$ has the weight $(w,n)=(1,1)$ and a special form of ${\cal F}$ giving ${\cal F}_{S\bar{S}}=0$, otherwise such a constraint does not appear. For the superconformal invariance, the superpotential $W$ should have $(3,3)$. From the weight condition, a possible form is $W=S^3g(\Phi^i)$ but this term is forbidden by the symmetry under $S\to Se^{i\Theta}$ which the D-term part $[{\cal F}]_D$ has. Therefore, we cannot add any superpotential terms of matter multiplets.
%%%%%%%%%%%%%%%%%%%%%%%%%%%%%%%%%%%%%%%%%%%%%%%%%%%%%%%%%%%%%%%%%%%%%%%%%%%%%%%%%%%%%%%%%%%%%%%%%%%%%%%%%%%%%%%%%%%%%%%%%%%%%%%%%%%
\section{Relation between our results and other works}\label{discussion}
Here, we comment on the differences between ours and the results in Ref.~\cite{Koehn:2012ar}, in which the DBI action of a chiral multiplet is constructed in the old minimal SUGRA. As we mentioned before, the DBI action of a real linear multiplet can be rewritten in terms of a chiral multiplet through the linear-chiral duality and the whole action of a chiral multiplet is obtained in global SUSY in terms of superfield \cite{GonzalezRey:1998kh}. The authors of Ref.~\cite{Koehn:2012ar} embedded the dual chiral multiplet action into the old minimal SUGRA. On the other hand, our starting point is the action of a real linear multiplet, more precisely, the constraint $\eqref{global constraint}$ imposed upon it. This constraint has its origin in the tensor multiplet of $\mathcal{N}=2$ SUSY \cite{Rocek:1997hi,Bagger:1997pi,GonzalezRey:1998kh}. Indeed, in global SUSY case, the real linear multiplet corresponds to a Goldstino multiplet for the broken SUSY. From such a viewpoint, our construction is important since it makes the connection with the partial breaking of $\mathcal{N}=2$ SUSY much clearer . 

Although the ways of construction are different, our action would realize their result. Indeed, at the bosonic component level, we have found the correspondence between the result in Ref.~\cite{Koehn:2012ar} and ours. However, we also found that the action cannot be realized in the old minimal SUGRA when we do not consider the case including higher-derivative terms of a chiral compensator, which may contradict the result of Ref.~\cite{Koehn:2012ar}. Unlike the DBI action of a real linear multiplet, that of a vector multiplet can be constructed in both of the old and new minimal SUGRA \cite{Abe:2015nxa}. The difference originates from the necessity of {\it u-associated} derivatives in the DBI action of a real linear multiplet. For a vector superfield case, we can construct the DBI action only with the chiral projection operator $\Sigma$, which does not require {\it u-associated} multiplet to make the operand superfield a primary superfield~\cite{Kugo:1983mv,Butter:2009cp,Kugo:2016zzf}. It is interesting to explore these reasons and we expect that the direct derivation of the constraint $\eqref{global constraint}$ and also DBI action from $\mathcal{N}=2$ SUGRA are necessary to understand this issue, which would be our future work \footnote{For the DBI action of a vector multiplet, such attempts have been recently discussed \cite{Kuzenko:2015rfx}. There, the partial breaking of ${\cal N}=2$ SUSY in some ${\cal N}=1$ SUSY background has been discussed.}.

%%%%%%%%%%%%%%%%%%%%%%%%%%%%%%%%%%%%%%%%%%%%%%%%%%%%%%%%%%%%%%%%%%%%%%%%%%%%%%%%%%%%%%%%%%%%%%%%%%%%%%%%%%%%%%%%%%%%%%%%%%%%%%%%%%%
\section{Summary}\label{summary}
In this paper, we have discussed superconformal generalization of a DBI action of a real linear superfield known in global SUSY. 

To achieve this, we have focused on the constraint $\eqref{global constraint}$ between a chiral multiplet and a real linear multiplet, which comes from the partial breaking of 4D $\mathcal{N}=2$ SUSY \cite{Bagger:1997pi}. However, it is a nontrivial task to embed this constraint into conformal SUGRA due to the existence of the SUSY spinor derivative, which in general, cannot be applied for arbitrary multiplets in conformal SUGRA. Instead of using an original spinor derivative, we have adopted the {\it{u-associated}} spinor derivative, proposed in Ref.~\cite{Kugo:1983mv}. We obtained the condition $\eqref{weight condition between w3n3 and w2n2}$ and $\eqref{explicit condotion for chiral}$ by requiring that the corresponding constraint $\eqref{u-derivative part}$ in conformal SUGRA becomes a chiral constraint. Surprisingly, we have found that these conditions can be realized only in the new minimal formulation of SUGRA when we choose the general power function of compensator as the {\it{u-associated}} multiplet. Then, we have derived the condition $\eqref{L0 condition2}$ which {\it{u-associated}} multiplets must satisfy.

After embedding the constraint into the new minimal SUGRA, we have shown the component action which is formulated in curved spacetime. We have also discussed the linear-chiral duality at the level of bosonic components and rewritten the action from a complex scalar field of a chiral multiplet. Finally, we have constructed the action where matter multiplets are directly coupled to the DBI sector. Due to the appearance of nonlinear terms for vector field $B_{0a}$, we have restricted the discussion to the special form of matter function $\eqref{special form}$ and derived the bosonic action $\eqref{special matter}$.  

In this paper, we have shown that the DBI action of a real linear multiplet cannot be realized in the old minimal SUGRA as a naive embedding of the constraint $\eqref{global constraint}$, which may contradict the result of Ref.~\cite{Koehn:2012ar}. The duality relation between the old and new minimal SUGRA \cite{Ferrara:1983dh} is generically not obvious when there exist higher-derivative terms. For example, the non-minimal coupling of gravity is realized only in new minimal SUGRA \cite{Farakos:2012je} as in the case of the DBI action we discussed here. Such an issue may be revealed with the help of deep understanding of SUGRA system with higher-order derivative terms.  

To investigate our model further, we need the direct derivation of the constraint from $\mathcal{N}=2$ SUGRA. And also, the remaining part in Eq. $\eqref{string DBI}$, i.e., a term including $B_{\mu \nu }$, and possible combinations of the Maxwell, scalar and 2-form parts have not been constructed. We leave them for future work.

%%%%%%%%%%%%%%%%%%%%%%%%%%%%%%%%%%%%%%%%%%%%%%%%%%%%%%%%%%%%%%%%%%%%%%%%%%%%%%%%%%%%%%%%%%%%%%%%%%%%%%%%%%%%%%%%%%%%%%%%%%%%%%%%%%%
\section*{Acknowledgment}
The authors would like to thank Taichiro Kugo for helpful discussions and comments. YY would like thank also to Hiroyuki Abe and Yutaka Sakamura for useful discussion and collaboration in the related work. The work of YY is supported by JSPS
Research Fellowships for Young Scientists No. 26-4236 in Japan.
%%%%%%%%%%%%%%%%%%%%%%%%%%%%%%%%%%%%%%%%%%%%%%%%%%%%%%%%%%%%%%%%%%%%%%%%%%%%%%%%%%%%%%%%%%%%%%%%%%%%%%%%%%%%%%%%%%%%%%%%%%%%%%%%%%%
\begin{appendix}
\section{The components of {\it{u-associated}} spinor derivative multiplet}\label{explicit}
Here we show the explicit component form of 
\begin{align}
\frac{1}{L_0}\bar{\mathcal{D}}^{(L_0)}L\bar{\mathcal{D}}^{(L_0)}L. \label{Lo derivative part2}
\end{align}
As we have seen in Sec.~\ref{extension}, Eq. $\eqref{Lo derivative part2}$ is a chiral multiplet with weight $(3,3)$. The components of this multiplet, $\{ z',P_L\chi' , F'\}$, are  
\begin{align}
z'&=\frac{C^2}{C_0} \left(\bar{\tilde{Z}}-\bar{\tilde{Z}}_0  \right) P_R\left(   \tilde{Z}-\tilde{Z}_0  \right),\\
\nonumber P_L\chi' &=\frac{\sqrt{2}C^2}{C_0}P_L\biggl[ \left(  \tilde{\slash {B}}-i\slash{D}\tilde{C}-\tilde{\slash {B}}_0+i\slash{D}\tilde{C}_0\right) \left( \tilde{Z}-\tilde{Z}_0 \right) -\frac{3i}{2}\tilde{Z}_0\bar{\tilde{Z}}_0P_R\tilde{Z}_0\\
&-\frac{i}{2}\tilde{Z}_0\bar{\tilde{Z}}P_R\tilde{Z}+\frac{i}{4}\gamma ^a\tilde{Z}_0\bar{\tilde{Z}}\gamma _a\gamma _5 \tilde{Z}+i\tilde{Z}\bar{\tilde{Z}}_0P_R\tilde{Z}_0-\frac{i}{2}\gamma ^a\tilde{Z}\bar{\tilde{Z}}_0\gamma _a\gamma _5 \tilde{Z}_0\biggr] ,\\
\nonumber F'&=\frac{C^2}{C_0}\biggl[ -\left( \tilde{B}_a-i D_a\tilde{C}\right) ^2 +2\left( \tilde{B}_a-i D_a\tilde{C} \right) \left( \tilde{B}^a-i D^a\tilde{C} \right) -\left( \tilde{B}_{0a}-i D_{0a}\tilde{C}\right) ^2\\
\nonumber &+i\bar{\tilde{Z}}_0\gamma _5\left(  \tilde{\slash {B}}-i\slash{D}\tilde{C} \right)\left( \tilde{Z}-\tilde{Z}_0  \right)+\frac{i}{2}\bar{\tilde{Z}} \gamma _5\left(  \tilde{\slash {B}}_0-i\slash{D}\tilde{C} _0\right)\tilde{Z}\\
\nonumber &-2i\bar{\tilde{Z}} \gamma _5\left(  \tilde{\slash {B}}_0-i\slash{D}\tilde{C} _0\right)\tilde{Z}_0+\frac{3i}{2}\bar{\tilde{Z}}_0 \gamma _5\left(  \tilde{\slash {B}}_0-i\slash{D}\tilde{C} _0\right)\tilde{Z}_0\\
\nonumber &+2\left( \bar{\tilde{Z}}-\bar{\tilde{Z}}_0 \right)P_R \slash{D}\left( \tilde{Z}-\tilde{Z}_0 \right) +\frac{1}{2}\bar{\tilde{Z}}_0P_R\tilde{Z}_0\bar{\tilde{Z}}\tilde{Z}+\frac{1}{2}\bar{\tilde{Z}}P_R\tilde{Z}\bar{\tilde{Z}}_0\tilde{Z}_0\\
&+2\bar{\tilde{Z}}P_R\tilde{Z}_0\bar{\tilde{Z}}\tilde{Z}_0-3\bar{\tilde{Z}}P_R\tilde{Z}_0\bar{\tilde{Z}}_0\tilde{Z}_0-3\bar{\tilde{Z}}\tilde{Z}_0\bar{\tilde{Z}}_0P_R\tilde{Z}_0+\frac{1}{2}\bar{\tilde{Z}}_0P_R\tilde{Z}_0\bar{\tilde{Z}}_0\tilde{Z}_0\biggr] ,
\end{align}
where the fields with $\tilde{}$ are divided by the first components of the multiplet they belong to, in the same way as Eq. $\eqref{tilde}$, and the superconformal derivative $D_a$ is understood to act only on the numerator but not on the denominator, e.g., $D^a\tilde{C} \equiv D^aC/C =D^a \log C$.

\end{appendix}


\begin{thebibliography}{99}


%\cite{Khoury:2010gb}
\bibitem{Khoury:2010gb} 
  J.~Khoury, J.~L.~Lehners and B.~Ovrut,
  ``Supersymmetric P(X,$\phi$) and the Ghost Condensate,''
  Phys.\ Rev.\ D {\bf 83}, 125031 (2011)
  %\cite{Khoury:2011da}
\bibitem{Khoury:2011da} 
  J.~Khoury, J.~L.~Lehners and B.~A.~Ovrut,
  ``Supersymmetric Galileons,''
  Phys.\ Rev.\ D {\bf 84}, 043521 (2011)
%\cite{Baumann:2011nm}
\bibitem{Baumann:2011nm} 
  D.~Baumann and D.~Green,
  ``Supergravity for Effective Theories,''
  JHEP {\bf 1203}, 001 (2012)
      %\cite{Farakos:2012je}
\bibitem{Farakos:2012je} 
  F.~Farakos, C.~Germani, A.~Kehagias and E.~N.~Saridakis,
  ``A New Class of Four-Dimensional N=1 Supergravity with Non-minimal Derivative Couplings,''
  JHEP {\bf 1205}, 050 (2012)
  %\cite{Koehn:2012ar}
\bibitem{Koehn:2012ar} 
  M.~Koehn, J.~L.~Lehners and B.~A.~Ovrut,
  ``Higher-Derivative Chiral Superfield Actions Coupled to N=1 Supergravity,''
  Phys.\ Rev.\ D {\bf 86}, 085019 (2012)
  %\cite{Farakos:2012qu}
\bibitem{Farakos:2012qu} 
  F.~Farakos and A.~Kehagias,
  ``Emerging Potentials in Higher-Derivative Gauged Chiral Models Coupled to N=1 Supergravity,''
  JHEP {\bf 1211}, 077 (2012)
  %\cite{Koehn:2012te}
\bibitem{Koehn:2012te} 
  M.~Koehn, J.~L.~Lehners and B.~Ovrut,
  ``Ghost condensate in $N=1$ supergravity,''
  Phys.\ Rev.\ D {\bf 87}, no. 6, 065022 (2013)
  %\cite{Farakos:2013zya}
\bibitem{Farakos:2013zya} 
  F.~Farakos, C.~Germani and A.~Kehagias,
  ``On ghost-free supersymmetric galileons,''
  JHEP {\bf 1311}, 045 (2013)
    %\cite{Gwyn:2014wna}
\bibitem{Gwyn:2014wna} 
  R.~Gwyn and J.~L.~Lehners,
  ``Non-Canonical Inflation in Supergravity,''
  JHEP {\bf 1405}, 050 (2014)
  %\cite{Aoki:2014pna}
\bibitem{Aoki:2014pna} 
  S.~Aoki and Y.~Yamada,
  ``Inflation in supergravity without Kahler potential,''
  Phys.\ Rev.\ D {\bf 90}, no. 12, 127701 (2014)
  %\cite{Aoki:2015eba}
\bibitem{Aoki:2015eba} 
  S.~Aoki and Y.~Yamada,
  ``Impacts of supersymmetric higher derivative terms on inflation models in supergravity,''
  JCAP {\bf 1507}, no. 07, 020 (2015)
  %\cite{Ciupke:2015msa}
\bibitem{Ciupke:2015msa} 
  D.~Ciupke, J.~Louis and A.~Westphal,
  ``Higher-Derivative Supergravity and Moduli Stabilization,''
  JHEP {\bf 1510}, 094 (2015)
  %\cite{Bielleman:2016grv}
\bibitem{Bielleman:2016grv} 
  S.~Bielleman, L.~E.~Ibanez, F.~G.~Pedro, I.~Valenzuela and C.~Wieck,
  ``The DBI Action, Higher-derivative Supergravity, and Flattening Inflaton Potentials,''
  arXiv:1602.00699 [hep-th].
  
  %\cite{Born:1934gh}
\bibitem{Born:1934gh} 
  M.~Born and L.~Infeld,
  ``Foundations of the new field theory,''
  Proc.\ Roy.\ Soc.\ Lond.\ A {\bf 144}, 425 (1934).
  %\cite{Dirac:1962iy}
\bibitem{Dirac:1962iy} 
  P.~A.~M.~Dirac,
  ``An Extensible model of the electron,''
  Proc.\ Roy.\ Soc.\ Lond.\ A {\bf 268}, 57 (1962).
  %%%%%%%%%%%%%%%%%%%%%%%%%%%%%%%%%%%%%%%%%%%%%%%%%%%%%%%%%%%%
  %\cite{Aganagic:1996pe}
\bibitem{Aganagic:1996pe} 
  M.~Aganagic, C.~Popescu and J.~H.~Schwarz,
  ``D-brane actions with local kappa symmetry,''
  Phys.\ Lett.\ B {\bf 393}, 311 (1997)
  %\cite{Aganagic:1996nn}
\bibitem{Aganagic:1996nn} 
  M.~Aganagic, C.~Popescu and J.~H.~Schwarz,
  ``Gauge invariant and gauge fixed D-brane actions,''
  Nucl.\ Phys.\ B {\bf 495}, 99 (1997)
  %\cite{Bergshoeff:2013pia}
\bibitem{Bergshoeff:2013pia} 
  E.~Bergshoeff, F.~Coomans, R.~Kallosh, C.~S.~Shahbazi and A.~Van Proeyen,
  ``Dirac-Born-Infeld-Volkov-Akulov and Deformation of Supersymmetry,''
  JHEP {\bf 1308}, 100 (2013)
%\cite{Howe:1996mx}
\bibitem{Howe:1996mx} 
  P.~S.~Howe and E.~Sezgin,
  ``Superbranes,''
  Phys.\ Lett.\ B {\bf 390}, 133 (1997)
  %\cite{Cederwall:1996pv}
\bibitem{Cederwall:1996pv} 
  M.~Cederwall, A.~von Gussich, B.~E.~W.~Nilsson and A.~Westerberg,
  ``The Dirichlet super three-brane in ten-dimensional type IIB supergravity,''
  Nucl.\ Phys.\ B {\bf 490}, 163 (1997)
  %\cite{Cederwall:1996ri}
\bibitem{Cederwall:1996ri} 
  M.~Cederwall, A.~von Gussich, B.~E.~W.~Nilsson, P.~Sundell and A.~Westerberg,
  ``The Dirichlet super p-branes in ten-dimensional type IIA and IIB supergravity,''
  Nucl.\ Phys.\ B {\bf 490}, 179 (1997)
  %\cite{Bergshoeff:1996tu}
\bibitem{Bergshoeff:1996tu} 
  E.~Bergshoeff and P.~K.~Townsend,
  ``Super D-branes,''
  Nucl.\ Phys.\ B {\bf 490}, 145 (1997)
  %%%%%%%%%%%%%%%%%%%%%%%%%%%%%%%%%%%%%%%%%%%%%%%%%%%%%%%%%%%%
  %\cite{Cecotti:1986gb}
\bibitem{Cecotti:1986gb} 
  S.~Cecotti and S.~Ferrara,
  ``Supersymmetric Born-infeld Lagrangians,''
  Phys.\ Lett.\ B {\bf 187}, 335 (1987).
  %\cite{Bagger:1996wp}
\bibitem{Bagger:1996wp} 
  J.~Bagger and A.~Galperin,
  ``A New Goldstone multiplet for partially broken supersymmetry,''
  Phys.\ Rev.\ D {\bf 55}, 1091 (1997)
%\cite{Rocek:1997hi}
\bibitem{Rocek:1997hi} 
  M.~Rocek and A.~A.~Tseytlin,
  ``Partial breaking of global D = 4 supersymmetry, constrained superfields, and three-brane actions,''
  Phys.\ Rev.\ D {\bf 59}, 106001 (1999)
  %\cite{Kuzenko:2002vk}
\bibitem{Kuzenko:2002vk} 
  S.~M.~Kuzenko and S.~A.~McCarthy,
  ``Nonlinear selfduality and supergravity,''
  JHEP {\bf 0302}, 038 (2003)
  %\cite{Kuzenko:2005wh}
\bibitem{Kuzenko:2005wh} 
  S.~M.~Kuzenko and S.~A.~McCarthy,
  ``On the component structure of N=1 supersymmetric nonlinear electrodynamics,''
  JHEP {\bf 0505}, 012 (2005)
  %\cite{Abe:2015nxa}
  \bibitem{Abe:2015nxa} 
  H.~Abe, Y.~Sakamura and Y.~Yamada,
  ``Matter coupled Dirac-Born-Infeld action in four-dimensional N=1 conformal supergravity,''
  Phys.\ Rev.\ D {\bf 92}, no. 2, 025017 (2015)
%\cite{Abe:2015fha}
\bibitem{Abe:2015fha} 
  H.~Abe, Y.~Sakamura and Y.~Yamada,
  ``Massive vector multiplet inflation with Dirac-Born-Infeld type action,''
  Phys.\ Rev.\ D {\bf 91}, no. 12, 125042 (2015)
   %\cite{Ferrara:2014oka}
\bibitem{Ferrara:2014oka} 
  S.~Ferrara, M.~Porrati and A.~Sagnotti,
  ``N = 2 Born-Infeld attractors,''
  JHEP {\bf 1412}, 065 (2014)
  %\cite{Ferrara:2014nwa}
\bibitem{Ferrara:2014nwa} 
  S.~Ferrara, M.~Porrati, A.~Sagnotti, R.~Stora and A.~Yeranyan,
  ``Generalized Born--Infeld Actions and Projective Cubic Curves,''
  Fortsch.\ Phys.\  {\bf 63}, 189 (2015)
  %\cite{Ferrara:2015ixa}
  
\bibitem{Ferrara:2015ixa} 
  S.~Ferrara and A.~Sagnotti,
  ``Massive Born--Infeld and Other Dual Pairs,''
  JHEP {\bf 1504}, 032 (2015)
  
  %\cite{Andrianopoli:2014mia}
\bibitem{Andrianopoli:2014mia} 
  L.~Andrianopoli, R.~D'Auria and M.~Trigiante,
  ``On the dualization of Born-Infeld theories,''
  Phys.\ Lett.\ B {\bf 744}, 225 (2015)
  %\cite{Andrianopoli:2015wqa}
\bibitem{Andrianopoli:2015wqa} 
  L.~Andrianopoli, R.~D'Auria, S.~Ferrara and M.~Trigiante,
  ``Observations on the partial breaking of $N=2$ rigid supersymmetry,''
  Phys.\ Lett.\ B {\bf 744}, 116 (2015)
  %\cite{Andrianopoli:2015rpa}
\bibitem{Andrianopoli:2015rpa} 
  L.~Andrianopoli, P.~Concha, R.~D'Auria, E.~Rodriguez and M.~Trigiante,
  ``Observations on BI from $\mathcal{N}=2$ Supergravity and the General Ward Identity,''
  JHEP {\bf 1511}, 061 (2015)
%\cite{Andrianopoli:2016eub}
\bibitem{Andrianopoli:2016eub} 
  L.~Andrianopoli, R.~D'Auria, S.~Ferrara and M.~Trigiante,
  ``C-map for Born-Infeld theories,''
  arXiv:1603.03338 [hep-th].
  

  
  %\cite{Bagger:1997pi}
\bibitem{Bagger:1997pi} 
  J.~Bagger and A.~Galperin,
  ``The Tensor Goldstone multiplet for partially broken supersymmetry,''
  Phys.\ Lett.\ B {\bf 412}, 296 (1997)
    %\cite{GonzalezRey:1998kh}
\bibitem{GonzalezRey:1998kh} 
  F.~Gonzalez-Rey, I.~Y.~Park and M.~Rocek,
  ``On dual 3-brane actions with partially broken N=2 supersymmetry,''
  Nucl.\ Phys.\ B {\bf 544}, 243 (1999)
  
    %\cite{Siegel:1979ai}
\bibitem{Siegel:1979ai} 
  W.~Siegel,
  ``Gauge Spinor Superfield as a Scalar Multiplet,''
  Phys.\ Lett.\ B {\bf 85}, 333 (1979).
  
  %\cite{Kugo:1978nz}
\bibitem{Kaku:1978nz} 
  M.~Kaku, P.~K.~Townsend and P.~van Nieuwenhuizen,
  ``Properties of Conformal Supergravity,''
  Phys.\ Rev.\ D {\bf 17}, 3179 (1978),
  
  M.~Kaku and P.~K.~Townsend,
  ``Poincare Supergravity As Broken Superconformal Gravity,''
  Phys.\ Lett.\ B {\bf 76}, 54 (1978),
  
  P.~K.~Townsend and P.~van Nieuwenhuizen,
  ``Simplifications of Conformal Supergravity,''
  Phys.\ Rev.\ D {\bf 19}, 3166 (1979).
  %\cite{Kugo:1982cu}
\bibitem{Kugo:1982cu} 
  T.~Kugo and S.~Uehara,
  ``Conformal and Poincare Tensor Calculi in $N=1$ Supergravity,''
  Nucl.\ Phys.\ B {\bf 226}, 49 (1983).
  %\cite{Kugo:1983mv}
\bibitem{Kugo:1983mv} 
  T.~Kugo and S.~Uehara,
  ``$N=1$ Superconformal Tensor Calculus: Multiplets With External Lorentz Indices and Spinor Derivative Operators,''
  Prog.\ Theor.\ Phys.\  {\bf 73}, 235 (1985).
    %\cite{Butter:2009cp}
\bibitem{Butter:2009cp} 
  D.~Butter,
  ``N=1 Conformal Superspace in Four Dimensions,''
  Annals Phys.\  {\bf 325}, 1026 (2010)
   %\cite{Kugo:2016zzf}
\bibitem{Kugo:2016zzf} 
  T.~Kugo, R.~Yokokura and K.~Yoshioka,
  ``Component versus Superspace Approaches to D=4, N=1 Conformal Supergravity,''
  arXiv:1602.04441 [hep-th].
  %\cite{Wess:1992cp}
\bibitem{Wess:1992cp} 
  J.~Wess and J.~Bagger, 1992,
  ``Supersymmetry and supergravity,'' Univ. Pr.. Princeton, USA: r. 259 p.
  %\cite{Freedman:2012zz}
\bibitem{Freedman:2012zz} 
  D.~Z.~Freedman and A.~Van Proeyen, 2012,
  ``Supergravity,'' Cambridge University Press

  %\cite{Volkov:1972jx}
\bibitem{Volkov:1972jx} 
  D.~V.~Volkov and V.~P.~Akulov,
  ``Possible universal neutrino interaction,''
  JETP Lett.\  {\bf 16}, 438 (1972)
  [Pisma Zh.\ Eksp.\ Teor.\ Fiz.\  {\bf 16}, 621 (1972)],
  
  D.~V.~Volkov and V.~P.~Akulov,
  ``Is the Neutrino a Goldstone Particle?,''
  Phys.\ Lett.\ B {\bf 46}, 109 (1973).
  
  %\cite{Rocek:1978nb}
\bibitem{Rocek:1978nb} 
  M.~Rocek,
  ``Linearizing the Volkov-Akulov Model,''
  Phys.\ Rev.\ Lett.\  {\bf 41}, 451 (1978),
  
  E.~A.~Ivanov and A.~A.~Kapustnikov,
  ``General Relationship Between Linear and Nonlinear Realizations of Supersymmetry,''
  J.\ Phys.\ A {\bf 11}, 2375 (1978),
  
  U.~Lindstrom and M.~Rocek,
  %``Constrained Local Superfields,''
  Phys.\ Rev.\ D {\bf 19}, 2300 (1979).
  

  
  %\cite{Ferrara:1983dh}
\bibitem{Ferrara:1983dh} 
  S.~Ferrara, L.~Girardello, T.~Kugo and A.~Van Proeyen,
  ``Relation Between Different Auxiliary Field Formulations of $N=1$ Supergravity Coupled to Matter,''
  Nucl.\ Phys.\ B {\bf 223}, 191 (1983).

  %\cite{Kuzenko:2015rfx}
\bibitem{Kuzenko:2015rfx} 
  S.~M.~Kuzenko and G.~Tartaglino-Mazzucchelli,
  ``Nilpotent chiral superfield in N=2 supergravity and partial rigid supersymmetry breaking,''
  arXiv:1512.01964 [hep-th].
  
\end{thebibliography}
\end{document}